\documentclass[preprint,showpacs,superscriptaddress,preprintnumbers,amsmath,amssymb]{revtex4-1}

\usepackage{amsfonts}
\usepackage{mathrsfs}
\usepackage{amsmath}
\usepackage[usenames,dvipsnames]{xcolor}
\usepackage{graphicx}
\usepackage{bm}
\usepackage{amssymb}
\usepackage{xspace}
\usepackage{epstopdf}
\usepackage{dcolumn}
\usepackage{longtable}
\usepackage{multirow}
\usepackage[colorlinks=true, letterpaper=true, pdfstartview=FitV, linkcolor=blue, citecolor=blue, urlcolor=blue]{hyperref}
\usepackage{romannum}

\usepackage{cancel}
\usepackage{ulem}
\newcommand{\jwadd}[1]{{\color{blue} #1}}

\begin{document}

	
\title{Origin of giant valley splitting in silicon quantum wells induced by superlattice barriers }
\author{Gang Wang}
\affiliation{ 
	State Key Laboratory of Superlattices and Microstructures, Institute of Semiconductors, Chinese Academy of Sciences, Beijing 100083, China}
\affiliation{ 
	Center of Materials Science and Optoelectronics Engineering, University of Chinese Academy of Sciences, Beijing 100049, China}
\author{Zhi-Gang Song}
\email{songzhigang@semi.ac.cn}
\affiliation{ 
	State Key Laboratory of Superlattices and Microstructures, Institute of Semiconductors, Chinese Academy of Sciences, Beijing 100083, China}
\author{Jun-Wei Luo}
\email{jwluo@semi.ac.cn}
\affiliation{ 
State Key Laboratory of Superlattices and Microstructures, Institute of Semiconductors, Chinese Academy of Sciences, Beijing 100083, China}
\affiliation{ 
Center of Materials Science and Optoelectronics Engineering, University of Chinese Academy of Sciences, Beijing 100049, China}
\affiliation{ 
Beijing Academy of Quantum Information Sciences, Beijing 100193, China}
\author{Shu-Shen Li}
\affiliation{ 
	State Key Laboratory of Superlattices and Microstructures, Institute of Semiconductors, Chinese Academy of Sciences, Beijing 100083, China}
\affiliation{ 
	Center of Materials Science and Optoelectronics Engineering, University of Chinese Academy of Sciences, Beijing 100049, China}

\date{\today}

\begin{abstract}
Enhancing valley splitting in SiGe heterostructures is a crucial task for developing silicon spin qubits. Complex SiGe heterostructures, sharing a common feature of four-monolayer (4ML) Ge layer next to the silicon quantum well (QW), have been computationally designed to have giant valley splitting approaching 9 meV. However, none of them has been fabricated may due to their complexity. Here, we remarkably simplify the original designed complex SiGe heterostructures by laying out the Si QW directly on the Ge substrate followed by capping a (Ge$_4$Si$_4$)$_n$ superlattice (SL) barrier with a small sacrifice on VS as it is reduced from a maximum of 8.7 meV to 5.2 meV.  Even the smallest number of periods ($n = 1$) will also give a sizable VS of 1.6 meV, which is large enough for developing stable spin qubits. We also develop an effective Hamiltonian model to reveal the underlying microscopic physics of enhanced valley splitting by (Ge$_4$Si$_4$)$_n$ SL barriers. We find that the presence of the SL barrier will reduce the VS instead of enhancing it. Only the (Ge$_4$Si$_4$)$_n$ SL barriers with an extremely strong coupling with Si QW valley states provide a remarkable enhancement in VS.
 These findings lay a solid theoretical foundation for the realization of sufficiently large VS for Si qubits.
  
\end{abstract}

\maketitle

\section{Introduction}
Silicon spin qubits have several advantages over other competing qubit schemes~\cite{zwanenburg_silicon_2013,ladd_silicon_2018} including significantly lengthened spin relaxation time~\cite{Tahan2014}, extremely long dephasing time~\cite{zwanenburg_silicon_2013}, and the mature microelectronic technologies~\cite{ladd_silicon_2018}.  Particularly, the last one is the key advantage for its capability to integrate both the qubits and electronics on a single chip to take advantage of both spin and charge degrees and fabricate millions of qubits in a silicon quantum chip needed to implement universal quantum computing. However, the two-fold degeneracy of the conduction band X-valleys becomes  a primary factor that hinders us to achieve the well-defined and effectively controlled silicon spin qubits~\cite{zwanenburg_silicon_2013,ando_electronic_1982,ladd_silicon_2018}. In bulk Si, the conduction band minimum (CBM) is located at $\Delta$-point [$0.85\times(2\pi/a_0)$ from the $\Gamma$-point toward the $X$-point of the Brillouin zone ($a_0=5.43$ \AA is the Si lattice constant] and thus has a sixfold valley degeneracy. In Si quantum wells (QWs), confinement along one direction lifts valley degeneracy, resulting in the six-fold degenerate $\Delta_6$ valleys splitting into a low-lying two-fold degenerate $\Delta_2$ valley pair along [001] direction, and a high-lying four-fold degenerate $\Delta_4$ valley.  Two-fold degeneracy of the low-lying $\Delta_2$ valleys is hard to further lift. However, if the ground-state manifold is composed of more than one orbital level, it is difficult to precisely manipulate the spin qubit and realize well-characterized qubits, which is a crucial criterion for the implementation of quantum computation~\cite{divincenzo_physical_2000}. Only if the magnetic field-induced qubit energy splitting (Zeeman splitting) is smaller concerning the valley splittings the latter contribution can be disregarded. In Si QWs, it was believed~\cite{ohkawa_theory_1977,boykin_valley_2004,nestoklon_spin_2006,friesen_valley_2007,valavanis_intervalley_2007,saraiva_physical_2009} that the sharp interfacial potential causes coupling between two low-lying valleys and thus  lifts the two-fold degeneracy with an energy separation referring to valley splitting (VS). Numerous studies~\cite{yang_spin-valley_2013,petit_spin_2018,koiller_exchange_2001,culcer_realizing_2009} have been carried out to evaluate the impacts of VS on the initialization and manipulation of Si spin qubits and to enhance VS. For instance, the magnitude of VS determines the highest possible operation temperature of the Si spin qubit, which is usually less than 30 mK~\cite{maurand_cmos_2016,huang_fidelity_2019,watson_programmable_2018} and was recently promoted to exceed 1 K~\cite{yang_operation_2020}. Spin relaxation hotspots (relaxation rate peak or rapid decline in spin lifetime) occur when the VS equals the Zeeman splitting, indicating a large VS is required to achieve a long spin lifetime~\cite{yang_spin-valley_2013,petit_spin_2018}. Last but not least, sizable VS is also essential for the fulfillment of two-electron qubits~\cite{klinovaja_exchange-based_2012,taylor_fault-tolerant_2005,levy_universal_2002}. Unfortunately, in Si qubit devices the achievable VS is remarkably limited~\cite{yang_spin-valley_2013,gamble_valley_2016,zajac_reconfigurable_2015,goswami_controllable_2006,wuetz_effect_2020,mi_high-resolution_2017,hollmann_large_2020,borselli_measurement_2011,borjans_single-spin_2019} to about $0.1\sim 0.3$ meV in MOS (Si/SiO$_2$) architecture~\cite{yang_spin-valley_2013,gamble_valley_2016} and to an even smaller value in Si/SiGe QW architecture ranging from 0.01 to 0.2 meV~\cite{zajac_reconfigurable_2015,goswami_controllable_2006,wuetz_effect_2020,mi_high-resolution_2017,hollmann_large_2020,borselli_measurement_2011,borjans_single-spin_2019}.

Previous work~\cite{zhang_genetic_2013} has computationally designed Si QWs with substantially enhanced VS approaching 9 meV relying on a combination of a genetic algorithm used to search optimizers in the configuration space and atomistic semi-empirical pseudopotential method (SEPM) for band structure calculations~\cite{folded_spectrum_Wang1994,SEPM_Wang1995,Wang1999}.  The searching configuration space is composed of a Si QW with a fixed thickness (say 40 ML, 1 ML = $a_0/4$, $a_0$ is the lattice constant) sandwiched between two mirror-symmetric 80-ML thick Ge/Si superlattices (SLs) with varying the stacking sequence of Ge/Si monolayers. For a 40-ML thick Si QW, the optimized best SL barriers are with stacking sequences of Ge$_4$Si$_4$Ge$_2$Si$_6$Ge$_4$Si$_4$Ge$_4$Si$_2$..., Ge$_4$Si$_4$Ge$_4$Si$_2$Ge$_4$Si$_6$Ge$_4$Si$_2$..., and Ge$_4$Si$_6$Ge$_2$Si$_6$Ge$_4$Si$_4$Ge$_4$Si$_4$... (subscript numbers denote the number of MLs) starting from the interface of 40-ML Si layer for substrates of pure Si and Si$_{0.8}$Ge$_{0.2}$ and  Si$_{0.6}$Ge$_{0.4}$ alloys, respectively. The corresponding VS is 5.7, 7.4 and 8.7 meV, respectively. Interestingly, all designed SL barriers share the same feature of a 4-ML Ge sublayer immediately next to the 40-ML Si QW layer. Even a simple (Ge$_4$Si$_4$)$_{10}$ SL (10 periods with a repeating unit of 4-ML Ge/4-ML Si) barrier gives rise to a remarkably large VS of $E_{\textrm{VS}}=7.2$ meV for 40-ML thick Si QW.  By contrast, all SL barriers starting from a non-4-ML Ge sublayer afford a much smaller VS for Si QWs~\cite{zhang_genetic_2013}. Although these computationally designed SL barriers offer one order of magnitude enhancement in VS for Si QWs, none of them has been fabricated to overcome the challenge due to valley splitting towards Si quantum computing. Their complex stacking sequence or a too large number of SL periods may cause difficulties in fabrication. It is thus highly desired to simplify these structures for the sake to make them readily accessible to experimentalists utilizing currently reliable technology. Particularly, a thinner SL barrier will benefit the improvement of gate voltage control capability since spin qubits in gate-defined Si quantum dots (QDs) are electrically insulated from top metal gates by a gate oxide insulating layer and a SiGe barrier layer.  It is also interesting to unravel the physics underlying the 4-ML Ge sublayer alone to achieve an order of magnitude enhancement in VS for Si QWs.

In this work, we aim to simplify the structures to make them feasible for fabrication and reveal the physics governing the VS enhancement by the 4-ML Ge sublayer for Si QWs. Specifically, we consider putting the Si QW directly on the Ge substrate followed by a capping layer of a Ge/Si SL (schematically shown in FIG.~\ref{fig:fig1}(a)) instead of sandwiching the Si QW between two mirror-symmetric Ge/Si SLs (schematically shown in FIG.~\ref{fig:fig1}(a)) in the original design~\cite{zhang_genetic_2013}. We name such simplified structures as Si$_{40}$ QW/(Ge$_m$Si$_m$)$_n$ SL barrier hybrid systems. By performing atomistic calculations, we examine VS of the Si$_{40}$ QW/(Ge$_m$Si$_m$)$_n$ SL barrier hybrid systems as reducing the number of SL periods $n$ for a variety of periodic thickness $m=2,3,4,5,6$. We indeed find that the replacement of two mirror-symmetric $m=4$ SLs by one side of $m=4$ SL degrades slightly the VS from 7.2 meV to 5 meV for a 40-ML thick Si QW.  Reducing the number of SL periods $n$ for simplification, VS decreases gradually. Fortunately, even at the smallest number of periods ($n=1$), it still has a sizable VS of 1.6 meV, which is large enough for developing stable spin qubits. Whereas, $m\neq 4$ SL barriers suppress the VS of Si QW to  less than 1 meV. To reveal the underlying microscopic physics, we expand the Hilbert space by additionally including the miniband-edge states of the SL barrier (SL-MBS) different from the traditional valley splitting theory, which considers only the lowest two valley states in the Si QW (QW-VS)~\cite{ohkawa_theory_1977,boykin_valley_2004,nestoklon_spin_2006,friesen_valley_2007,valavanis_intervalley_2007,saraiva_physical_2009}. Based on this consideration, we establish the effective Hamiltonian model of the hybrid system.  In the model, the coupling between QW-VS and SL-MBS states is treated as off-diagonal elements.  After employing the L\"{o}wding partitioning method,  we obtain the reduced Hamiltonian for the two lowest valley states with the effects of SL-MBS treated as first-order perturbation~\cite{winkler_spin-orbit_2003}. It is thus straightforward to get VS of the hybrid system in a simple but intuitive formula, providing a deep insight into the significant enhancement of VS by the SL barrier.  We find that the presence of the SL barrier will reduce the VS instead of enhancing it. Only the $m=4$ SL barrier with an extremely strong coupling with Si QW valley states provides a remarkable enhancement in VS.

\section{Theoretical methods}
\subsection{Atomistic calculation method} 
We investigate here the Si QW and GeSi SL interaction effects on the valley coupling between the two lowest Si valley states using a direct diagonalization of the band Hamiltonian $-{1\over
2}\nabla^2+V({\bf r})$ for the QW/SL hybrid structure described by its potential $V({\bf r})$. We use
for the potential of the hybrid structure a superposition of overlapping,
spherical screened pseudopotential $v_{\alpha}(r)$ of the
constituent atom \cite{SEPM_Wang1995, Wang1999},
\begin{equation}
V({\bf r})=\sum_n\sum_{\alpha}\hat{v}_{\alpha}({\bf r}-{\bf R}_{n}-{\bf
	d}_{\alpha}),
\end{equation}
where $\hat{v}_{\alpha}({\bf r}-{\bf R}_{n}-{\bf d}_{\alpha})$ is
the screened pseudopotential containing spin-orbit interaction of
atom type $\alpha$ at site ${\bf d}_{\alpha}$ in the $n$th primary cell
${\bf R}_n$. Considering the 4.2\% lattice mismatch between Si and Ge, a large strain exists between atomic thin Si and Ge layers in GeSi SL. We use the atomistic valence force field (VFF) method \cite{Pryor1998,Williamson2000} to minimize the strain energy finding the atomic equilibrium positions. We diagonalize $-{1\over 2}\nabla^2+V({\bf r})$ within a
plane-wave basis \cite{SEPM_Wang1995} whose size is sufficiently selected such that the weak valley coupling is accurately considered. The supercell approach combined with the periodic boundary conditions is implemented. The screened pseudopotentials
$\{\hat{v}_{\alpha}\}$ are fitted \cite{SEPM_Wang1995} so as to remove
the "LDA error" in the bulk crystal; they reproduce well not only
the band gaps throughout the zone, but also the electron and hole
effective-mass tensors, as well as the valence band and conduction
band offsets between well and barrier materials, spin-orbit
splittings, and GW spin-splitting in bulk materials\cite{JWL_bulk}.
This is described in Ref. \onlinecite{SEPM_Wang1995, JWL_bulk}. This
approach has been previously applied to superlattices \cite{Magri_PRB, Wang_PRL97}, colloidal quantum dots \cite{An_NL},
and Stranski-Krastanow (SK) quantum dots \cite{GB_NP}.

\subsection{Effective Hamiltonian model}
The hybrid system consists of  [001]-oriented Si QW and (Ge$_{m}$Si$_{m}$)$_n$ SL barrier. These two components possess $D_{2d}$, $C_{2v}$ point symmetries. The whole system  is of $C_1$ symmetry in the QW growth direction. Symmetry allows the couplings between the QW-VS and the SL-MBS because these states share the single representation of the point group (in terms of the double group, including spin-orbit coupling). These couplings may influence the VS of the Si QW subunit.  To access this influence,  we expand the Hilbert space of the lowest two Si valley states~\cite{ohkawa_theory_1977,boykin_valley_2004,nestoklon_spin_2006,friesen_valley_2007,valavanis_intervalley_2007,saraiva_physical_2009} to include SL-MBS of the SL barrier and have the effective Hamiltonian for the hybrid system by introducing the coupling between QW-VS and SL-MBS states (off-diagonal elements) as follows:   
\begin{equation}\label{eq1}
H=\left[
\begin{array}{ccccccc}
\epsilon_{0}^{\textrm{QW}} & \delta  &\lambda_{1}^{1} & \text{...} & \lambda_{i}^{1} & \text{...} & \lambda_{n}^{1}
\\
\delta  & \epsilon_{0}^{\textrm{QW}} & \lambda_{1}^{2} & \text{...} &\lambda_{i}^{2} & \text{...} & \lambda_{n}^{2}
\\
\lambda_{1}^{1} & \lambda_{1}^{2} & \epsilon_{1}^{\textrm{SL}} & 0 & 0 & 0 & 0 \\
\text{...} & \text{...} & 0 & \text{...} & 0 & 0 & 0 \\
\lambda_{i}^{1} & \lambda_{i}^{2} & 0 & 0 & \epsilon_{i}^{\textrm{SL}} & 0 & 0 \\
\text{...} & \text{...} & 0 & 0 & 0 & \text{...} & 0 \\
\lambda_{n}^{1} & \lambda_{n}^{2} & 0 & 0 & 0 & 0 & \epsilon_{2n}^{\textrm{SL}}%
\end{array}%
\right]
\end{equation}
where the diagonal element $\epsilon_{0}^{\textrm{QW}}$ is the average of  the lowest two degenerate valley states in the Si QW. This two-fold degeneracy is lifted in the energy of $2\delta$ by the confinement potential. Therefore, the energy levels of the two Si valleys are $\epsilon_{1}^{\textrm{QW}}=\epsilon_{0}^{\textrm{QW}}-\delta$, and $\epsilon_{2}^{\textrm{QW}}=\epsilon_{0}^{\textrm{QW}}+\delta$, respectively. $\epsilon_i^{SL}$ is the energy level of $i$th state in (Ge$_m$Si$_m$)$_n$ SL ($i=1,2,...,2n$). Note that each period has  $\Delta_z$ and $\Delta_{-z}$  two valley states, and thus SL with n periods has $2n$ valley states. The remaining $\Delta_x$ and $\Delta_y$ valley states are neglected due to weak coupling between $\Delta_z$ and $\Delta_{x,y}$. $\lambda_{i}^{1,2}$ depict the interaction strengths between two Si valley states and the $i^{th}$ state of the (Ge$_m$Si$_m$)$_n$ SL barrier. Figs.~\ref{fig:fig2}(a)\,(b) schematically illustrate these SL-MBS and QW-VS as well as their couplings.

The substantial space confinement in the short-period SL barrier yields energy levels of the SL-MBS much higher than that of the QW-VS, with an energy difference being far more extensive than $\delta$. We, therefore, can reduce the $(2+2n)$-dimension effective Hamiltonian Eq.\,(\ref{eq1}) to a $2\times2$ Hamiltonian for the lowest two Si valley states taking  into account the effects of SL-MBS as first-order perturbation based on the quasi-degenerate perturbation theory using the L\"{o}wding partitioning method~\cite{winkler_spin-orbit_2003}. The reduced effective Hamiltonian is thus as follows:
\begin{equation}\label{eq2}
\tilde{H}=\left[
\begin{array}{cc}
\tilde{H}_{11} & \tilde{H}_{12} \\
\tilde{H}_{21} & \tilde{H}_{22}%
\end{array}%
\right]
\end{equation}
where,
\begin{subequations}
	\begin{eqnarray}\label{eq3}
	\tilde{H}_{11} &=&\epsilon_{0}^{\textrm{QW}}+\sum_{i}^{2n}\frac{%
		\lambda_{i}^{1}\lambda_{i}^{1}}{\epsilon_{0}^{\textrm{QW}}-\epsilon_{i}^{\textrm{SL}}}, \\
	\tilde{H}_{22} &=&\epsilon_{0}^{\textrm{QW}}+\sum_{i}^{2n}\frac{\lambda_{i}^{2}\lambda_{i}^{2}}{\epsilon_{0}^{\textrm{QW}}-\epsilon_{i}^{\textrm{SL}}}, \\
	\tilde{H}_{12} &=&\tilde{H}_{21}=\delta +\sum_{i}^{2n}\frac{\lambda_{i}^{1}\lambda_{i}^{2}}{%
		\epsilon_{0}^{\textrm{QW}}-\epsilon_{i}^{\textrm{SL}}}.
	\end{eqnarray}
\end{subequations}

The reduced Hamiltonian can now be  diagonalized directly, and  the corresponding eigenvalues of two valley states read,
\begin{equation}\label{eq4}
 E_{\pm} = \frac{1}{2}\left(\tilde{H}_{11}+\tilde{H}_{22}\pm E_{\text{VS}}\right),
\end{equation}
and the VS of the hybrid system is,
\begin{equation}\label{eq5}
E_{\text{VS}} = \sqrt{(\tilde{H}_{11}-\tilde{H}_{22})^{2}+4\tilde{H}_{12}%
	\tilde{H}_{21}}.
\end{equation}
If we learn $\lambda_i^{1,2}$,  $\epsilon_{0}^{\textrm{QW}}$, and $\epsilon_{i}^{\textrm{SL}}$, we can then reproduce $E_{\textrm{VS}}$ predicted from atomistic calculations according to Eq.\,(\ref{eq5}), which will tell the factors' contribution to the enhancement of  VS due to coupling between QW-VS and SL-MBS ($\lambda_i^{1,2}$).

\section{Computational results}
\begin{figure}[!hbp]
	\includegraphics[width=0.8\columnwidth]{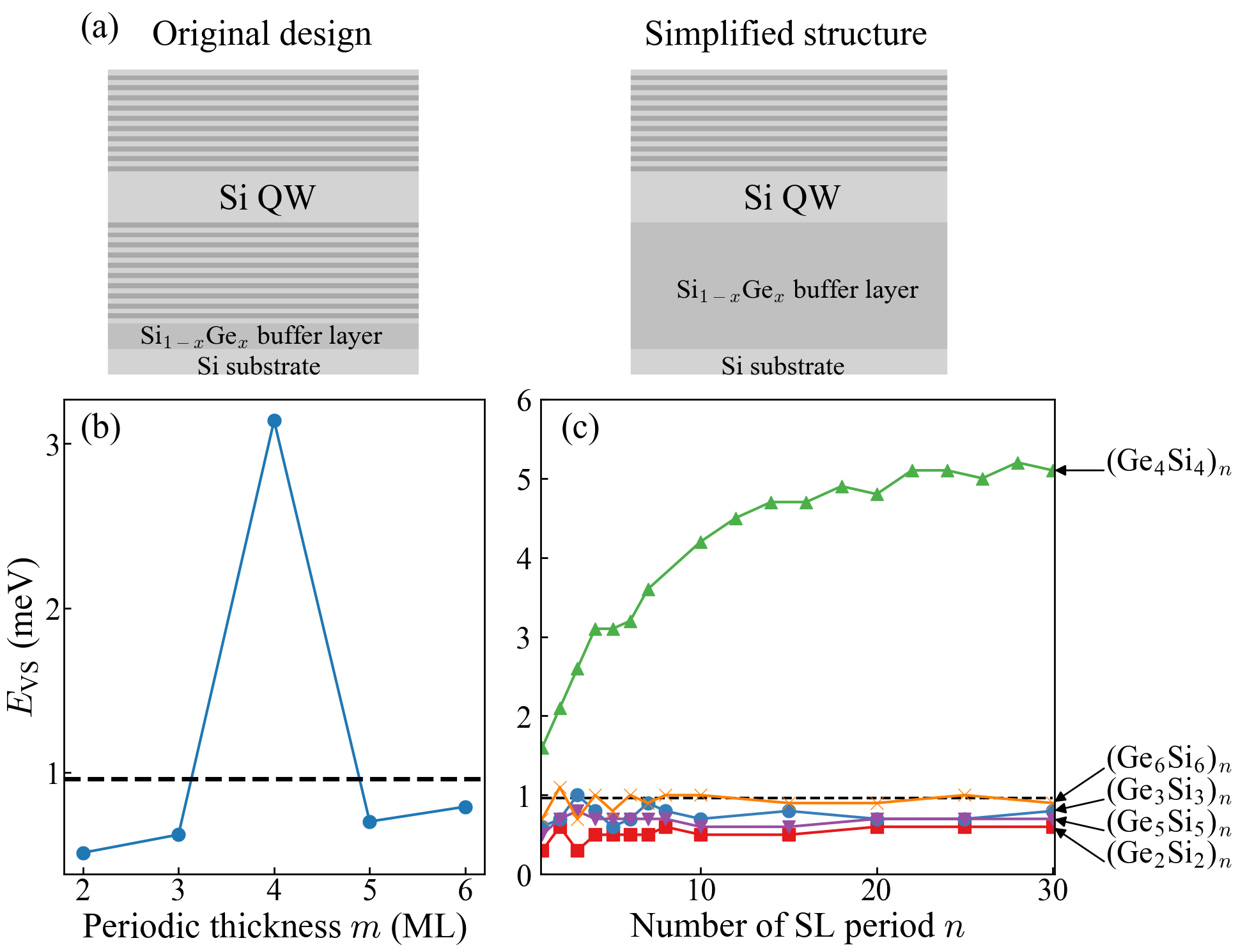} 
	\caption{\label{fig:fig1} (a) Schematic sketch of Si QW structure in the original design (left) and simplified Si QW structure in this work (right). (b) Computationally predicted VS of Si$_{40}$ QW/(Ge$_{m}$Si$_{m}$)$_{4}$ SL barrier hybrid systems as varying period thickness $m$ from 2 to 6 by performing atomistic SEMP calculations.  The dashed horizon line marks the VS $E_{\textrm{VS}}$ of the corresponding isolated 40-ML Si QW embedded in the pure Ge barrier.
   (c) Computationally predicted $E_{\textrm{VS}}$ of Si$_{40}$ QW/(Ge$_{m}$Si$_{m}$)$_{4}$ SL barrier hybrid systems as a function of the number of SL periods $n$ for period thickness $m=2,3,4,5,6$.}
\end{figure}

FIG.~\ref{fig:fig1}(c) shows atomistic pseudopotential method calculated VS $E_{\textrm{VS}}$  of the Si$_{40}$ QW/(Ge$_{m}$Si$_{m}$)$_{n}$ SL barrier hybrid systems as a function of the number of SL periods $n$ for a variety of period thickness $m=2,3,4,5,6$. The VS of the isolated 40-ML Si QW embedded in the pure Ge barrier is \jwadd{0.9 meV} (represented by the dashed lines in Figs. 1(b) and (c)). Interestingly, inserting just one unit of Ge$_4$Si$_4$ (i.e., $n=1$ in the case of  $m=4$ SL barrier) into the interface between Si QW and Ge barrier, the VS is immediately enhanced to $E_{\textrm{VS}}=1.6$ meV. This means we can considerably simplify the hybrid system by reducing the number of (Ge$_{4}$Si$_{4}$)$_{n}$ SL periods to $n=1$ to have a large enough VS for Si electron spin qubits~\cite{yang_spin-valley_2013}. Increasing the number of (Ge$_{4}$Si$_{4}$)$_{n}$ SL periods from $n=1$ to $n=10$, $E_{\textrm{VS}}$ raises rapidly from 1.6 meV to 4.2 meV. After that, $E_{\textrm{VS}}$ grows slowly toward a saturation value of 5 meV as further increasing $n$. We note that this saturation value is only slightly smaller than $E_{\textrm{VS}}=7.2$ meV of the original sandwich structure of the (Si)$_{40}$ QW/(Ge$_{m}$Si$_{m}$)$_\text{10}$ SL barrier~\cite{zhang_genetic_2013} despite, the former being much simpler than the latter for experimental fabrication as shown in FIG.~\ref{fig:fig1}(a). Whereas, in cases of $m\neq 4$ SL barriers, we see that $E_{\textrm{VS}}$ is even smaller than the VS ($E_{\textrm{VS}}=0.9$ meV) of the isolated Si QW in a whole range of investigated periods $n=1-30$. This result is unexpected. It also implies that the enhanced $E_{\textrm{VS}}$ by the $m=4$ SL barrier is susceptible to the fluctuation of the atomic thickness in SL repeating units, which requires precisely controlled growth of Si/Ge interfaces at the atomic layer level.

\begin{figure}[!hbp]
	\includegraphics[width=0.6\columnwidth]{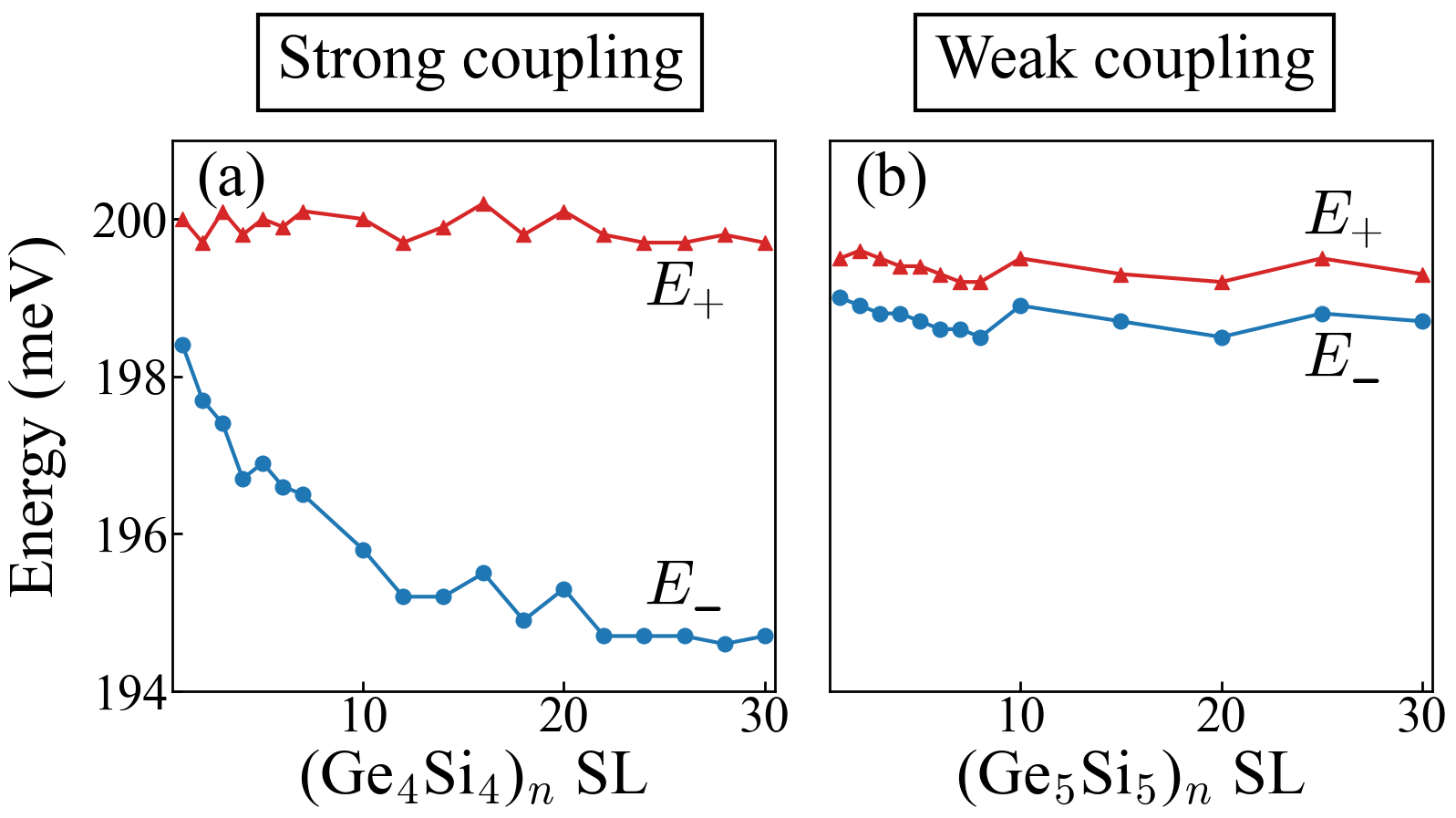} 
	\caption{\label{fig:fig3} Atomistic calculations predicted energy levels of the two lowest Si valley states in (a) Si$_{40}$ QW/(Ge$_4$Si$_4$)$_n$ SL barrier and (b) Si$_{40}$ QW/(Ge$_5$Si$_5$)$_n$ SL barrier as a function of the number of SL periods $n$.}
\end{figure}

We also examine the corresponding energy levels of the two lowest valley states of the hybrid systems as a function of the number of SL periods $n$ for $m=4$ and $m=5$ SL barriers. FIG.~\ref{fig:fig3}(a) shows the results for Si QW/(Ge$_4$Si$_4$)$_n$ SL barrier hybrid system. It is interesting to see that, as increasing the number of SL periods $n$, the upper level stays almost constant with a small fluctuation and the lower one goes down in energy, leading to the increase in $E_{\textrm{VS}}$, as shown in FIG.~\ref{fig:fig1}(c). Whereas, in the case of (Ge$_5$Si$_5$)$_n$ SL barrier, FIG.~\ref{fig:fig3}(b) shows that both energy levels are insensitive to varying the number of SL periods $n$, yielding a period-independent VS (see FIG.~\ref{fig:fig1}(c)).

\section{Discussion}
\subsection{The effect of the SL barrier states on VS}
To understand the computationally calculated results as mentioned above, we have developed an effective Hamiltonian model as presented in the method section. However, it requires assessing $\lambda_i^{1,2}$,  $\epsilon_{0}^{\textrm{QW}}$, and $\epsilon_{i}^{\textrm{SL}}$, which are difficult to obtain. In the following, we take further approximations. Firstly, we expect the coupling strengths between each SL state and two Si valley states to be approximately the same, \textit{i.e.}, $\lambda^{1}_{i}\sim \lambda^{2}_{i}$, regarding two Si valley states have a similar envelop wavefunction (see FIG.~\ref{fig:fig4}(a)) and  $2\delta \ll \epsilon_{i}^{SL}-\epsilon_{0}^{QW}$. For the sake of simplicity, we make a rough assumption that all SL-MBS have the same coupling strength to the QW-VS, \textit{i.e.}, $\lambda^{1}_{i}\sim \lambda^{2}_{i}\sim \Lambda$.
As a result, we can further simplify Eq.\,(\ref{eq4}) and Eq.\,(\ref{eq5}) as,
\begin{subequations}
	\begin{eqnarray} \label{eq6a}
	E_{\pm} &=&\epsilon_{0}^{\textrm{QW}}-\Lambda^{2}T\pm \frac{1}{2}E_{\text{VS}},  \\ \label{eq6b}
	E_{\text{VS}} &=&2\left\vert \Lambda^{2}T-\delta \right\vert,
	\end{eqnarray}
\end{subequations}
where 
\begin{equation}\label{eq7}
T \equiv \sum_{i}^{2n}\frac{1}{\epsilon_{i}^{\textrm{SL}}-\epsilon_{0}^{\textrm{QW}}}.
\end{equation}
Where $\Lambda$ quantifies the averaged coupling strength between SL-MBS and QW-VS, and $T$ is an energy factor representing the energy distribution of the SL-MBS relative to QW-VS. Therefore, $\Lambda^{2}T$ quantifies the total QW/SL coupling strength.

\begin{figure}[!hbp]
	\includegraphics[width=0.8\columnwidth]{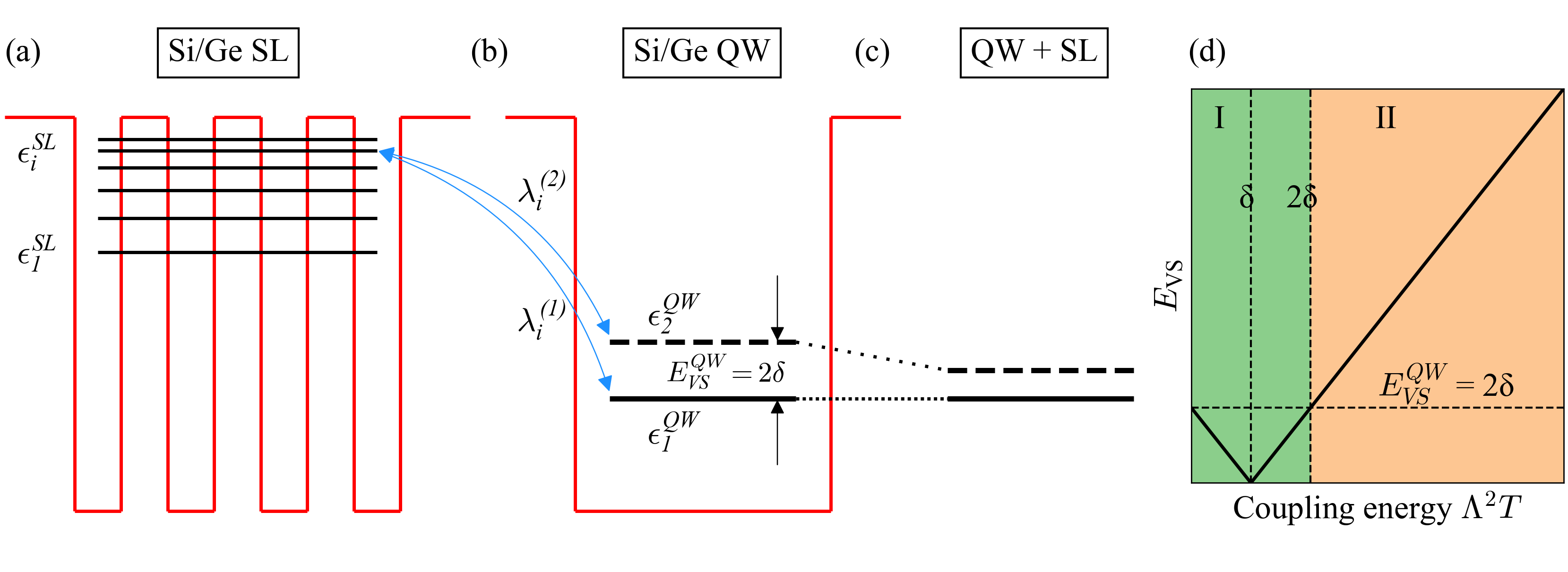} 
	\caption{\label{fig:fig2} 
		(a-b) Schematic representation of the interaction between the QW-VS and the SL-MBS. (c) Energies of the two Si valley states due to the QW/SL coupling. (d) VS as the function of the coupling energy; \Romannum{1} is the weak coupling region, and \Romannum{2} is the strong coupling region. Only when the coupling energy exceeds the critical value, namely $\Lambda^{2}T>2\delta$, the corresponding VS starts to be enhanced.}
\end{figure}

It is straightforward to read from Eq.\,(\ref{eq6a}) that, if $\Lambda^{2}T \leqslant  \delta$, $E_+=\epsilon_{0}^{\textrm{QW}}+\delta-2\Lambda^2T$ and  $E_-=\epsilon_{0}^{\textrm{QW}}-\delta$, otherwise $E_+=\epsilon_{0}^{\textrm{QW}}-\delta$ and $E_-=\epsilon_{0}^{\textrm{QW}}+\delta-2\Lambda^2T$. Interestingly, we can learn that the QW/SL coupling will not alter the energy level of the lower one of the two valley states in Si QW but push down the upper one. This model result explains exactly the above observation in the atomistic calculations (FIG.~\ref{fig:fig3}) that there is one energy level that stays almost constant with a small fluctuation as varying the number of periods $n$ as well as the period thickness $m$. It has also been schematically illustrated in Fig.~\ref{fig:fig2}(c). As increasing the QW/SL strength  $\Lambda^2T$ from zero to $\delta$, the upper valley approaches toward the lower valley, resulting in the reduction in $E_{\textrm{VS}}$. At $\Lambda^{2}T =  \delta$,  the upper valley passes the lower valley, eliminating the VS ($E_{\textrm{VS}}=0$). Afterward, further increasing the QW/SL strength  $\Lambda^2T$, the original upper valley (now becomes lower in energy) continuously goes down in energy, raising again the VS, but being still smaller than $2\delta$ the VS of the isolated Si QW. The enhancement in VS (i.e., $E_{\textrm{VS}}>2\delta$)  occurs only when $\Lambda^2T>2\delta$. Consequently, the presence of the SL barrier is, surprisingly, suppressing the VS instead of enhancing it unless the QW/SL coupling strength is strong enough.This finding explains why the VS of hybrid systems for all $m\neq4$ SL barriers is even smaller than that of the isolated Si QW, as shown in FIG.~\ref{fig:fig1}(c).

\begin{figure}[!hbp]
	\includegraphics[width=0.6\columnwidth]{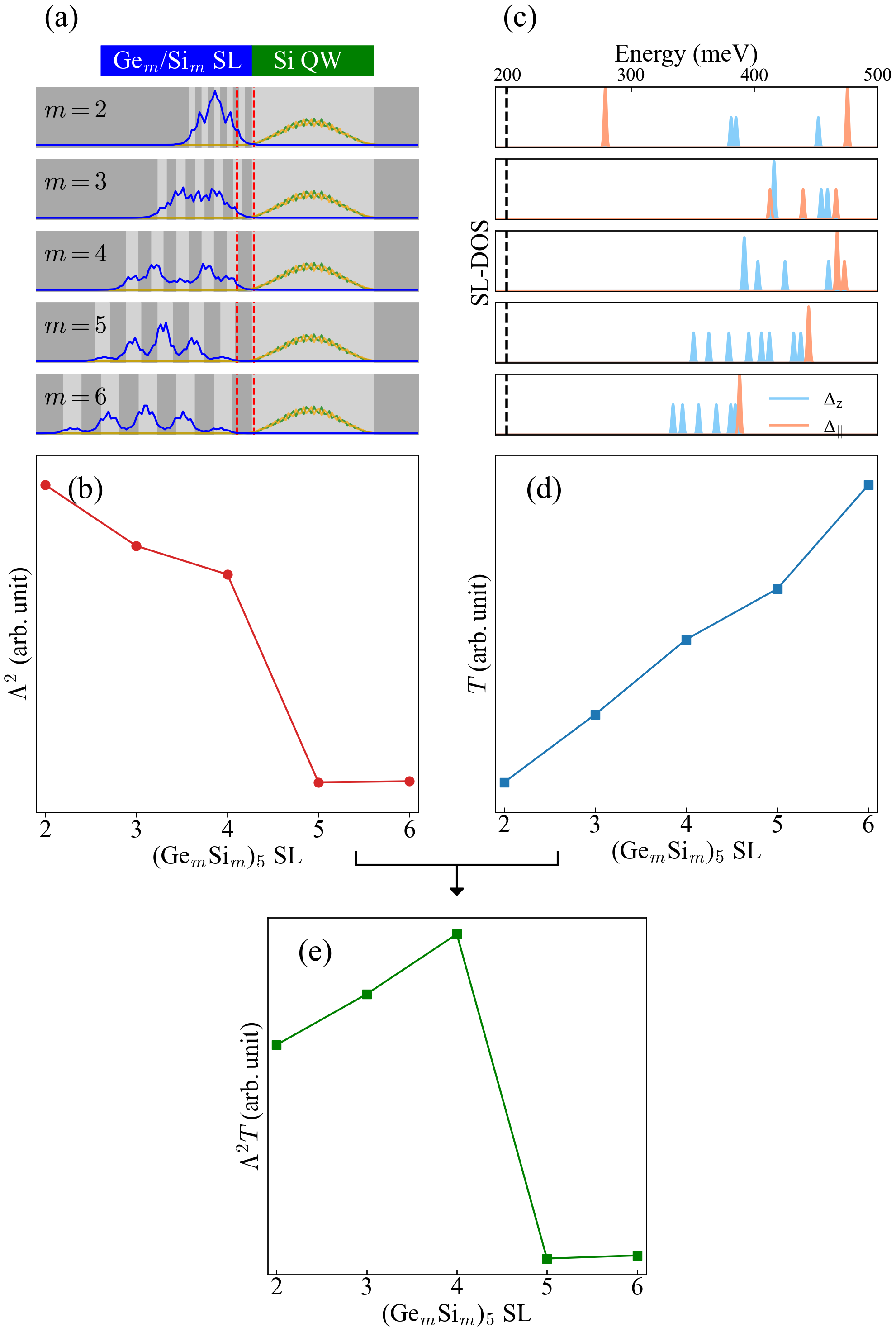} 
	\caption{\label{fig:fig4} (a) Wavefunction overlap between the QW-VS and the SL-MBS around the interface. The red dashed lines mark their overlap region. (b) The estimated coupling matrixes for different Ge$_m$Si$_m$ SL. (c) Energy spectrum of different Ge$_m$Si$_m$ SL (with SL period of 5). The reciprocal-space characters are quantified using the 'majority representation' approach~\cite{Wang1998} combined with the weight functions~\cite{Luo2008}. The black dashed lines mark the position of Si valley states in isolated Si QW. (d) The calculated energy factors $T$ for different Ge$_m$Si$_m$ SL. (e) The total coupling strength $\Lambda^{2}T$ between different Ge$_m$Si$_m$ SL and Si QW. }
\end{figure}

\begin{figure}[!hbp]
	\includegraphics[width=1\columnwidth]{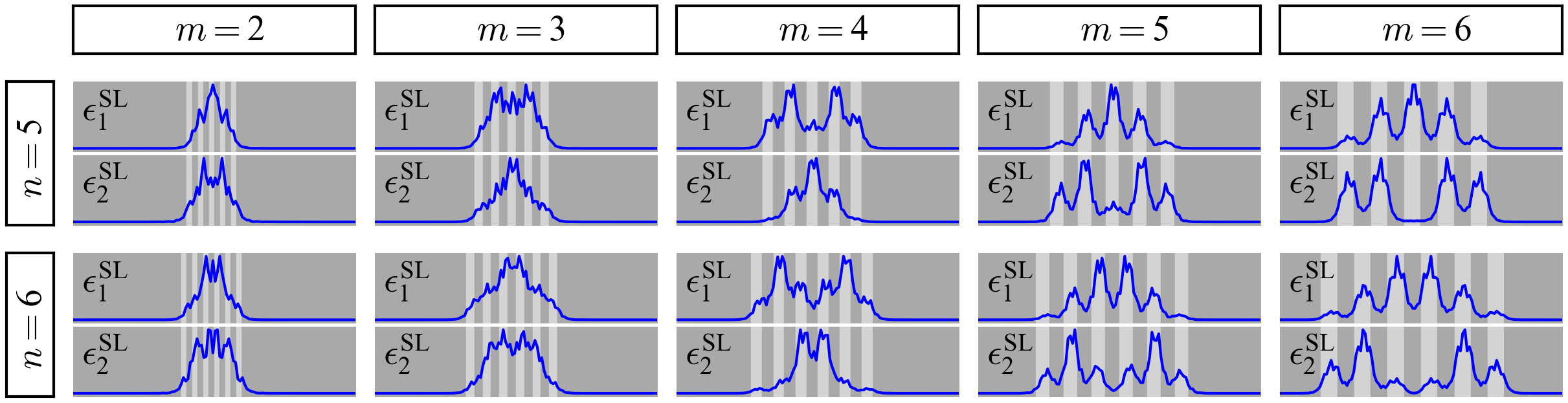} 
	\caption{\label{fig:fig5} Atomistic calculated wave functions of two lowest SL-MBS in (Ge$_m$Si$_m$)$_n$ SL with variant $m$ and $n$. }
\end{figure}

Fig.~\ref{fig:fig2}(d) sketches that we can divide the QW/SL coupling into two regions: strong coupling ($\Lambda^{2}T>2\delta$) and weak coupling ($\Lambda^{2}T\leqslant 2\delta$). From the results presented in FIG.~\ref{fig:fig1}(b)(c) we learn that the (Ge$_4$Si$_4$)$_n$ SL barrier provides a strong QW/SL coupling and the remaining  $m\neq 4$ (Ge$_m$Si$_m$)$_n$ SL barriers are in the weak coupling region. In the following, we attempt to unravel the physics causing the $m=4$ SL barrier to stand out clearly from the remaining $m\neq 4$ SL barriers by examining the coupling matrix $\Lambda$ and energy factor $T$ separately. 

\subsection{The coupling matrix $\Lambda$}
Following the definition of the coupling matrix $\Lambda \sim \left\langle \psi^{\textrm{QW}}|\Delta V|\psi^{\textrm{SL}}\right\rangle$, we can thus assess its magnitude by examining the wave function overlaps around the interfaces, as shown in Fig.~\ref{fig:fig4}(a). Ref.~\onlinecite{zhang_genetic_2013} has shown that the SL barriers always make wave functions of the lowest two valley states strongly localized inside the Si QW layer with a much smaller leakage compared with the SiGe alloy barrier. However, there is no one-to-one relationship between VS and wave function leakage. The difference in (Ge$_m$Si$_m$)$_n$ SL barriers introduces hardly a sizable change in wave functions of two valley states, as shown in Fig.~\ref{fig:fig4}(a). Here, we go stead to examine the penetration of the high-lying states of the SL barriers into the Si QW layer. We roughly approximate the coupling matrix $\Lambda$ as the evanescent integral of the (Ge$_m$Si$_m$)$_n$ SL wave functions: $\Lambda \sim \int_{z_0}^{z_1}\left| \Psi_{i}^{\textrm{SL}}(z)\right| ^{2}dz$ (the integral is running from $z_0$ inside the SL barrier  to $z_1$ inside the Si QW, indicated by red dashed lines in FIG.~\ref{fig:fig4}(a)). The evanescent integrals for different (Ge$_m$Si$_m$)$_5$ SL barriers are given in Fig.~\ref{fig:fig4}(b). One can see that the magnitude of $\Lambda$ decreases as the SL period thickness $m$ increases. Note that, in SLs made by semiconductors, the electron states  in neighboring QWs can interact as the barrier width decreases to a sufficient narrow thickness. The corresponding discrete energy levels confined inside QWs broaden into energy bands known as minibands, which are very narrow in energy compared to bulk energy bands~\cite{yu_fundamentals_2010}.   In the Si/Ge heterostructures, it is well known that the electron states are localized inside the Si layer~\cite{Mayeul2012}. Thus, thicker Ge sublayers of the (Ge$_m$Si$_m$)$_n$ SL cause SL miniband states to be less expanded and creates narrower minibands.  Furthermore, the thicker Si$_m$ sublayer will also give the electron states more space confinement, making their wave functions more localized inside Si sublayers. Fig.~\ref{fig:fig4} (a) indeed shows that the wave function in (Ge$_6$Si$_6$)$_n$ SL is the most localized while the wave function in (Ge$_2$Si$_2$)$_n$ SL is the most extended. Therefore, electron states of SL with larger $m$ have less probability of tunneling into the Si QW region. In this respect, a smaller coupling matrix $\Lambda$ is expected for (Ge$_m$Si$_m$)$_n$ SL with larger $m$, as shown in Fig.~\ref{fig:fig4}(b) based on a rough evaluation from the evanescent integral.

\subsection{The energy factor $T$.} 
We now turn to evaluate the dependence of the energy factor $T$ on the period thickness $m$ of the (Ge$_m$Si$_m$)$_n$ SL barrier in hybrid systems. To do so, we have also computed the unperturbed energy levels $\epsilon_{0}^{\textrm{QW}}$ and $\epsilon_{i}^{\textrm{SL}}$ of isolated Si QW and five-period (Ge$_m$Si$_m$)$_5$ SL, respectively, by carrying out atomistic SEPM calculations. Both isolated Si QW and five-period (Ge$_m$Si$_m$)$_5$ SL are separately embedded within the pure Ge matrix. FIG.~\ref{fig:fig4}(c) shows the atomistic calculations predicted results. One can see that, as the period thickness is getting thinner (or $m$ decreases), the SL energy levels $\epsilon_{i}^{\textrm{SL}}$ raise and are going far away from the Si QW valley level $\epsilon_{0}^{\textrm{QW}}$. This raising in the SL energy levels is attributed to the enhanced quantum confinement effect. The energy spacing between SL energy levels $\epsilon_{i}^{\textrm{SL}}$ is also getting larger, responsible for a wider miniband width.  We sum up the SL energy levels relative to $\epsilon_{0}^{\textrm{QW}}$ according to Eq.\,(\ref{eq7}), giving rise to the energy factor $T$, which is shown in FIG.~\ref{fig:fig4}(d) as a function of SL period thickness $m$. One can see that $T$ gets larger almost linearly as increasing the SL period thickness $m$. It is straightforward to learn that this linear increase in $T$ is due to going down in energy of all $\epsilon_{i}^{\textrm{SL}}$ relative to $\epsilon_{0}^{\textrm{QW}}$. 

Furthermore, if the number of SL periods $n$ is infinite, the discrete energy levels in the SL form minibands. The summation in $T$ term can then be approximated by the integral over a fixed energy interval from miniband bottom ($\epsilon_{1}^{\textrm{SL}}$) to miniband top ($\epsilon_{2n}^{\textrm{SL}}$): $T \equiv \sum_{i}\frac{1}{\epsilon_{i}^{\textrm{SL}}-\epsilon_{0}^{\textrm{QW}}}\propto\int_{\epsilon_{1}^{\textrm{SL}}}^{\epsilon_{2n}^{\textrm{SL}}}\frac{1}{\epsilon_{i}^{\textrm{SL}}-\epsilon_{0}^{\textrm{QW}}}d\epsilon=\ln(\frac{\epsilon_{2n}^{\textrm{SL}}-\epsilon_{0}^{\textrm{QW}}}{\epsilon_{1}^{\textrm{SL}}-\epsilon_{0}^{\textrm{QW}}})$. This approximation tells that, in addition to the absolute energy level positions, the miniband width is also an important parameter. Wider miniband width $\epsilon_{2n}^{\textrm{SL}}-\epsilon_{1}^{\textrm{SL}}$ occurred in smaller period thickness will give larger $T$ and thus larger VS, which compensates partially the reduction in T and thus VS induced by energy level shifting up due to strong confinement in narrow period thickness.

\subsection{Non-monotonic dependence of coupling strength $\Lambda^2 T$ on SL period thickness $m$.} 
So far, we have shown that by decreasing the periodic thickness of the SL barriers but keeping the number of periods constant, the coupling matrix $\Lambda^2$ is going to become bigger, and  the energy factor $T$ will get smaller for hybrid systems. FIG.~\ref{fig:fig4}(e) shows that the opposite trends in $\Lambda^2$ and $T$ render their product the total coupling strength $\Lambda^2T$ to be a non-monotonic function as varying the periodic thickness $m$ with a sharp peak occurring at $m=4$. This result is in excellent agreement with the atomistic calculation predicted  $E_{\textrm{VS}}$ for the hybrid systems, as shown in FIG.~\ref{fig:fig1}(a). Therefore, we have illustrated that the competition between the coupling matrix and energy factor makes $m=4$ SL barrier defeating the remaining $m\neq 4$ SL barriers to possess alone the strong coupling. In contrast, the hybrid systems associated with the remaining $m\neq 4$ SL barriers are in the weak coupling region, with $E_{\textrm{VS}}$ being smaller than the VS of the isolated Si QW, which is 0.96 meV for an isolated 40-ML Si QW embedded in the pure Ge matrix.

\subsection{Physics underlying the improvement of VS by $m=4$ SL barrier} 
We can now explain why only the (Ge$_4$Si$_4$)$_n$ SL barrier will continuously enhance the VS by increasing the number of SL periods $n$; whereas the remaining $m\neq 4$ SL barriers lack such enhancement, as shown in FIG.~\ref{fig:fig1}(c). Because the number of SL periods quantifies the total number of quantum states in the miniband~\cite{gerald_wave_1990},  the number of the valley states composing the SL miniband grows doubly as increasing the number of SL periods $n$, which has been explained in the text above Eq.\,(\ref{eq1}). According to Eq.\,(\ref{eq7}), a more significant number of SL periods $n$ gives a more considerable energy factor $T$.  On the other hand, increasing the SL periods  is expected to reduce the component of wavefunctions penetration into the Si QW, which will suppress the coupling matrix $\Lambda$. FIG.~\ref{fig:fig5} shows the spatial  distributions of the two lowest SL-MBS in (Ge$_m$Si$_m$)$_n$ SL with varying $m$ and $n$. One can observe a unique feature of the $m=4$ SL: the lowest energy ground state is $p$-like as it has one node in its envelope function. Whereas, remaining $m\neq4$ SLs have an  $s$-like ground state as usual.  FIG.~\ref{fig:fig5} exhibits that the p-like ground state in the $m=4$ SL guarantees more distributions of wavefunctions at the two terminated interfaces. That, in turn, ensures the component penetration into the Si QW, equally the coupling matrix $\Lambda$, is unchanged as increasing $n$.
In contrast,  the $s$-like ground state in the remaining $m\neq 4$ SLs is distributed mainly inside the SL with tails on the two terminated interfaces. In these $m\neq 4$ SLs, increasing the number of SL periods will increase the component inside the SL and thus reduce the component penetration into the Si QW. Subsequently, as the SL period increases, the coupling matrix $\Lambda$ decreases for $m\neq4$ SLs. The rise of $T$ and decrease of $\Lambda$ may make the $E_{\textrm{VS}}$ a product of $\Lambda$ and $T$ unchanged in $m\neq 4$ SLs, as shown in FIG.~\ref{fig:fig1}(c). However, the increased $T$ and unchanged $\Lambda$ make the $E_{\textrm{VS}}$ enhanced continuously as increasing $n$ in $m=4$ SL.

\section{Conclusion}
To make fabrication feasible, we have remarkably simplified the originally designed structure of the Si QW sandwiched by two symmetric Ge/Si SLs with substantially enhanced valley splitting $E_{\textrm{VS}}$ exceeding 9 meV~\cite{zhang_genetic_2013}. The simplified structure is engineered by laying out the Si QW on the Ge substrate and then capping a (Ge$_4$Si$_4$)$_n$ SL barrier on top of the Si QW. By performing the sophisticated atomistic-pseudopotential-calculations, we predicted that such simplification leads to a small sacrifice on $E_{\textrm{VS}}$, reducing it from the original 8.7 meV to 5.2 meV. We reduce the SL period $n$ in the (Ge$_4$Si$_4$)$_n$ barrier to further simplify the structure. In that case, $E_{\textrm{VS}}$ decreases gradually but still has a large enough value of $E_{\textrm{VS}}=1.6$ meV needed for electron spin qubits. Interestingly, the $m=4$ SL barrier is unique regarding  $E_{\textrm{VS}}$ is less than 1 meV  in structures associated with $m\neq 4$ SL barriers. To reveal the underlying microscopic physics, we have developed an effective Hamiltonian model of the simplified structure and thus obtained $E_{\textrm{VS}}$  in a simple but intuitive formula, providing insight into the enhancement of VS by SL barrier.  Surprisingly, we found that the presence of the SL barrier is usually suppressing the $E_{\textrm{VS}}$  instead of enhancing unless the QW/SL coupling strength is strong enough. We demonstrated that only the (Ge$_4$Si$_4$)$_n$ SL barrier has a strong coupling between QW and SL and thus gives rise to significantly enhanced $E_{\textrm{VS}}$. In contrast, the remaining $m\neq 4$ SL barriers have weak coupling and thus yield $E_{\textrm{VS}}$ even smaller than the corresponding isolated Si QW value. Our results provide a novel but effective, more importantly, easily fabricated approach to fulfill the large VS in Si that makes Si spin qubit gain new momentum in pursuing the general quantum computation.

\begin{acknowledgments}
The work was supported by the National Science
Fund for Distinguished Young Scholars under Grant
No. 11925407, the Basic Science Center Program of the National
Natural Science Foundation of China (NSFC) under
Grant No. 61888102, and the Key Research Program of Frontier
Sciences, CAS under Grant No. ZDBS-LY-JSC019.
\end{acknowledgments}

\nocite{*}
\bibliography{Ref_VSorigin}

\end{document}